\documentclass[%
 reprint,
 amsmath,amssymb,
 aps,
]{revtex4-1}

\usepackage[dvipsnames]{xcolor}

\usepackage{tikz} 
\usetikzlibrary{shapes.geometric}

\usepackage{graphicx}
\usepackage{dcolumn}
\usepackage{bm}

\begin{document}

\preprint{APS/123-QED}

\title{Monte Carlo Tensor Network Renormalization}

\author{William Huggins$^{1}$,  C. Daniel Freeman$^{1}$, Miles Stoudenmire$^{2}$, Norm M. Tubman$^{1*}$, K. Birgitta Whaley$^{1}$}
\affiliation{$^{1}$University of California, Berkeley, Berkeley, CA 94720, USA\\
$^{2}$Center for Computational Quantum Physics, Flatiron Institute, 162 5th Avenue, New York, NY 10010, USA\\
*Corresponding author:  ntubman@berkeley.edu
 }
\date{\today}

\begin{abstract}
Techniques for approximately contracting tensor networks are limited in how efficiently they can make use of parallel computing resources. In this work we demonstrate and characterize a Monte Carlo approach to the tensor network renormalization group method which can be used straightforwardly on modern computing architectures. We demonstrate the efficiency of the technique and show that Monte Carlo tensor network renormalization provides an attractive path to improving the accuracy of a wide class of challenging computations while also providing useful estimates of uncertainty and a statistical guarantee of unbiased results.
\end{abstract}

\maketitle

\section{\label{sec:level1}Introduction}

In the decades since the invention of the density matrix renormalization group~\cite{White1992} (DMRG) algorithm for determining the ground state of quantum systems, great strides have been made in understanding and generalizing its success. These developments includes many applications to 1D and 2D problems as well as small molecular chemistry Hamiltonians~\cite{dmrg2011,white1993,white1999,yanai2009,stoudenmire2012,Changlani2016,amaya2015}. The realization that DMRG could be seen as an efficient algorithm for variationally optimizing over a one dimensional matrix product state ansatz~\cite{Dukelsky1998} led to the development of tensor network wave functions, including projected entangled pair states (PEPS)~\cite{Verstraete2008} and the multiscale entanglement renormalization ansatz~\cite{Vidal2008}. By tailoring the graph structure and constraints in the constituent tensors, these new ansatz have been designed specifically to tackle higher dimensional spaces and ground states of systems near critical points. Other efforts (which we might call ``calculational methods'') have eschewed the variational approach entirely, instead directly representing the partition functions of classical or quantum systems as tensor networks and seeking efficient schemes for their approximate contraction~\cite{Levin2007, Xie2012Coarse-grainingDecomposition, Evenbly2015}.

Methods in this second class, as well as variational calculations for two dimensional systems with the PEPS ansatz, all face a common problem. Namely, that they involve tensor network contractions whose cost naively scales exponentially with system size. Fortunately, a host of algorithms, such as the tensor renormalization group~\cite{Levin2007} (TRG), tensor network renormalization~\cite{Evenbly2015}, and the corner transfer matrix renormalization group~\cite{Nishino1996CornerMethod}, have been developed to overcome this naive exponential cost with well-controlled approximations. These advancements have led to cutting-edge PEPS calculations of two dimensional fermionic lattice systems~\cite{Corboz2010SimulationStates,Corboz2016ImprovedModel,Zheng2016StripeModel}, extremely accurate results for thermodynamic properties of challenging classical Hamiltonians~\cite{Xie2012Coarse-grainingDecomposition}, and new strategies for directly accessing critical exponents~\cite{Evenbly2015}.

Despite these promising results, severe challenges still hamper the broad program of creating tensor network methods that match DMRG's power beyond the domain of one dimensional quantum systems~\cite{stoudenmire2012,bruognolo2017,simeng2010} and it is these challenges that motivate our exploration of stochastic tensor network contraction techniques. PEPS calculations for many Hamiltonians in two dimensions are computationally taxing, and, in three dimensions, mostly beyond the reach of current algorithms~\cite{Orus2012ExploringSystems}. Methods that aim to directly calculate the properties of a quantum system by contracting the tensor network representation of a partition function are more severely afflicted by the same difficulties. The inclusion of an imaginary time dimension means that applying these ``calculational methods'' even to two dimensional quantum systems requires the approximate contraction of tensor networks with a three dimensional structure. As a result, only simple contraction schemes carried out with small bond dimensions are currently feasible~\cite{Xie2012Coarse-grainingDecomposition}.

Tensor network methods have already benefited from the tremendous growth of computing power in recent years, and there have been several proposals for taking greater advantage of modern computing architectures, even proposals aimed at quantum computing~\cite{Schwarz2017ApproximatingStates}. Strikingly, however, these algorithms have not yet taken full advantage of the power of distributed high performance computing, although there has been some work in this regard~\cite{Chan2004AnCalculations, stoudenmire2013}. Making use of modern supercomputers by parallelizing the basic operations of tensor network calculations across hundreds or thousands of compute nodes promises to be a difficult feat likely to require an effort hand-tailored to the particular algorithm in question. In contrast, many Monte Carlo techniques used to simulate quantum systems use sampling techniques in which minimal communication is needed to communicate information between samples~\cite{needs2010,tubman2011,tubman2014,brown2013,zhang2003} and straightforward scaling to large numbers of nodes can be achieved easily. 

In this paper, we will investigate how effectively the recently proposed tensor network Monte Carlo~\cite{Ferris2015}, a flexible and naively parallelizable perfect sampling scheme for the stochastic evaluation of tensor networks, can be used to push tensor network renormalization group schemes beyond the current state of the art. We begin in section~\ref{sec:tnmc} by reviewing the sampling approach of Ref.~\cite{Ferris2015} in the language of the singular value decomposition. We explain in section \ref{sec:mcrg} how this method can be combined with a broad class of tensor renormalization algorithms while retaining its strong guarantees about unbiased results at any bond dimension and point out some of the particular choices that we have made in our implementation. In order to study the impact of these stochastic approximations in a well-understood setting we present the results of benchmark calculations on the 2d classical Ising model in section~\ref{sec:ising}. Finally, we argue that these results point towards an effective strategy for ameliorating the difficulties that state of the art algorithms encounter in approximately contracting higher dimensional tensor networks and in making effective use of high performance computing resources.

\section{\label{sec:tnmc}Stochastic Truncation with the Singular Value Decomposition}
The use of the singular value decomposition (SVD) to generate a low-rank tensor approximation is a key component of many tensor network algorithms. For concreteness, consider a tensor $T$ of order $k$: $T_{i_1, i_2... i_k}$. After dividing the indices of $T$ into two sets, $i_1, i_2... i_j$, and $i_{j+1}, i_{j+2}... i_k$, $T$ can be treated as 
a matrix $T_{m,n}$ indexed by a tuple \(m=(i_1, i_2,..., i_j)\) of elements from the first set and a tuple \(n=(i_{j+1}, i_{j+2},..., i_k)\) of elements from the second set. Taking the SVD of this matrix, we have
\begin{align}
	\label{eq:exactSVD}
	T_{mn} = \sum^{\chi}_{i} U_{mi} S_{ii} V_{in}
\end{align}

where \(U\) and \(V\) are unitary matrices and \(S\) is a rectangular diagonal matrix whose diagonal elements, the singular values, are non-negative real numbers. The best rank $\tilde{\chi}$ approximation (in the sense of minimizing the Frobenius norm of the difference between the exact $T$ and the approximation $\tilde{T}$)  is then given by discarding all but the $\tilde{\chi}$ largest elements of the diagonal matrix $S$~\cite{Eckart1936TheRank}, yielding:
\begin{align}
	\label{eq:approxSVD}
	T_{mn} \approx \tilde{T}_{mn} = \sum_{i}^{\tilde{\chi}} U_{mi} S_{ii} V_{in}
\end{align}
It was recently shown that one can sample from an ensemble of rank \(\tilde{\chi}\) approximations of $T$ (for any \(\tilde{\chi} < \chi\)) and exactly recover $T$ as the average of an infinite number of samples~\cite{Ferris2015}.

In the following section we explain how to perform this sampling. Allow $\mathcal{E}$ to denote an ensemble of samples, with a particular element $\mathbf{e}$ of this ensemble defined by a subset of size \(\tilde{\chi}\) of the nonzero singular values of the untruncated matrix \(T\),
\begin{equation}
\mathbf{e} = \{ s_{1},...,s_{\tilde{\chi}}\},\: s_i \in \{{S_{11},...,S_{\chi\chi}\}}.
\end{equation}
Let the matrix \(S^{(\mathbf{e})}\) be a diagonal matrix with the same shape as \(S\) but having \(\tilde{\chi}\) nonzero entries, determined from \(\mathbf{e}\) in a way which we will specify later. Then we define \(\tilde{T}^{(\mathbf{e})} = U S^{(\mathbf{e})} V\), a rank \(\tilde{\chi}\) matrix which could also be equivalently but more compactly expressed terms of submatrices of \(S^{(\mathbf{e})}\), \(U\), and \(V\).

In this approach we demand that the collection of rank $\tilde{\chi}$ matrices $\tilde{T}^{(\mathbf{e})}_j$ satisfy:
\begin{align}
	\label{eq:Tsum}
	\lim_{N \Rightarrow \infty} \frac{1}{N} \sum_{j}^{N} \tilde{T}^{(\mathbf{e})}_j = T
\end{align}
By substituting the definitions of \(T\) and the \(\tilde{T}^{(\mathbf{e})}\) into Eq.~\ref{eq:Tsum} we see that \(U\) and \(V\) can be canceled, yielding
\begin{align}
	\label{eq:Ssum}
	\lim_{N \Rightarrow \infty} \frac{1}{N} \sum_{j}^{N} S^{(\mathbf{e})}_j = S.
\end{align}
Or, in other words, the matrices $S^{(\mathbf{e})}_j$ must average to $S$. Understanding this, we can formulate a constructive procedure for generating the matrices \(S^{(\mathbf{e})}\). 
We begin with the original \(S\) and randomly select a subset of the singular values $\mathbf{e}$ (with probability $p(\mathbf{e})$) to keep, setting the rest to zero. In order to satisfy Eq.~\ref{eq:Ssum} we then rescale the retained singular values. We do this by multiplying each of them by the inverse of the probability of including that particular value in an individual sample: \(\frac{1}{r(S_{ii})}\). We note that this works for a general set of inclusion rates $r(S_{ii})$, determined from the subset selection probabilities \(p(\mathbf{e})\) by the following expression,
\begin{align}
r(S_{ii}) = \sum_{\mathbf{e}:\  S_{ii} \in \mathbf{e}} p(\mathbf{e}).
\end{align}

To illustrate, we consider a diagonal matrix element of Eq. \ref{eq:Ssum}: 
as N goes to infinity we average a sequence of terms which are equal to $\frac{S_{ii}}{r(S_{ii})}$ with probability $r(S_{ii})$ and are otherwise zero, resulting in eventual convergence to the value $S_{ii}$. In this way, after the reweighting, any choice of scheme for selecting the subsets that has a finite probability of including each nonzero singular value $S_{ii}$ will cause Eq. \ref{eq:Ssum}, and hence Eq. \ref{eq:Tsum}, to be satisfied and therefore lead to a valid ensemble. Guided by a desire to minimize the expectation value of the error, $||T - \tilde{T}^{(e)}||^2$, we set the relative probability of each sample $\mathbf{e}$ to
\begin{align}
w(\mathbf{e}) = \prod_{j}^{\tilde{\chi}} (s_{j})^{2}
\end{align}
and normalize these weights to form the probability distribution
\begin{align}
p(\mathbf{e}) = \frac{w(\mathbf{e})}{\sum_\mathbf{n} w(\mathbf{n})}.
\end{align}

We refer the reader to Ref.~\cite{Ferris2015} for details on a method to efficiently sample from this distribution and determine the probabilities of selection. We note that this can be performed with a time complexity $\mathcal{O}(\chi\tilde{\chi})$ where $\chi$ is the number of diagonal elements of $S$ and $\tilde{\chi}$ is the number of elements in the sample. With this sampling scheme in hand, we turn to its application as a component of an algorithm for tensor network renormalization.
\section{\label{sec:mcrg}Monte Carlo Renormalization}
\begin{figure}
	\begin{tikzpicture}[darkstyle/.style={circle,draw,fill=gray!40,minimum size=10}]
 		\foreach \x in {0,...,7}
	   	\foreach \y in {0,...,7} {
	      	\node [darkstyle]  (\x;\y) at (\x,\y) {};
	    }
        
	 	\foreach \x in {0,...,6}
	   	\foreach \y in {0,...,7} {
        	\pgfmathtruncatemacro\xp{\x+1}
            \draw (\x;\y)--(\xp;\y) node [midway, fill=white, scale=.5] (\x;\y;1) {\(\uparrow, \downarrow\)};
        }
        
	 	\foreach \x in {0,...,7}
	   	\foreach \y in {0,...,6} {
        	\pgfmathtruncatemacro\yp{\y+1}
            \draw (\x;\y)--(\x;\yp) node [midway, fill=white, scale=.5] (\x;\y;2) {\(\uparrow, \downarrow\)};
        }
        
	 	\foreach \x in {0,...,6}
	   	\foreach \y in {0,...,6} {
        	\pgfmathtruncatemacro\xp{\x+1}
        	\pgfmathtruncatemacro\yp{\y+1}
            \draw [red] (\x;\y;1)--(\x;\y;2);
            \draw [red] (\x;\y;1)--(\xp;\y;2);
            \draw [red] (\x;\yp;1)--(\x;\y;2);
            \draw [red] (\x;\yp;1)--(\xp;\y;2);
        }
    \end{tikzpicture}
    \caption{An 8x8 tensor network for the partition function of a 128 spin Ising model. The spin variables live on the bonds of the tensor network and neighboring spins have been connected by red lines to illustrate the relationship between the geometry of the lattice model and of the tensor network. Each tensor contains the interaction terms between the spins surrounding it. Bonds crossing the periodic boundary are omitted for clarity's sake. See Ref.~\cite{Evenbly2015} for more details.}
    \label{fig:partition_function_network}
\end{figure}
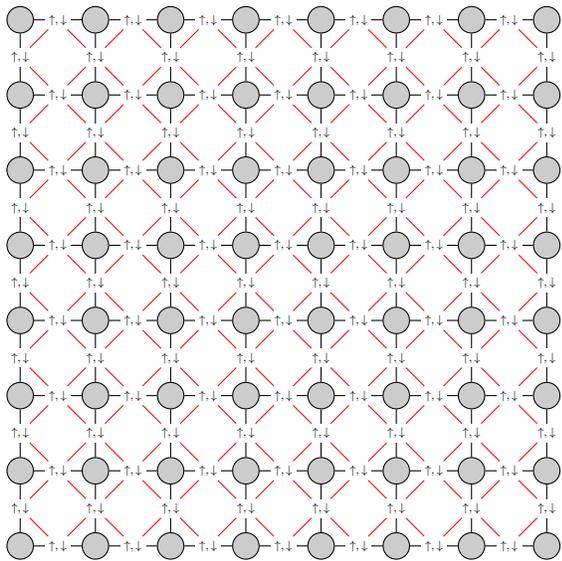
Let $\mathcal{N}_I$ be a tensor network, composed of tensors \(T_0, T_1, ..., T_k\), whose (tensor) trace represents some quantity of interest. We will specifically consider networks whose geometry is amenable to contraction with the tensor renormalization group (TRG) algorithm~\cite{Levin2007}, an iterative procedure for approximately contracting a tensor network that usually employs the singular value decomposition based truncation of Eq. ~\ref{eq:approxSVD}. However, as long as the truncation steps can be reduced to an application of the singular value decomposition, the following applies generally to other schemes for contraction by renormalization. As a concrete example of such a tensor network, we could take the partition function of a classical lattice model with local interactions on a square grid. We will consider here the case where all of the indices of $\mathcal{N}_I$ are summed over but the same arguments hold also when calculating a partial trace. Note that the notation in this section is somewhat different from the last section and that we use Roman numeral subscripts to refer to different levels of coarse-graining.

For each tensor $T_i \in \mathcal{N}_I$ which we must approximate by a truncated singular value decomposition, we can define an appropriate ensemble $\mathcal{E}_i$ as above such that $T_i = \langle \tilde{T}_i^{(\mathbf{e}_i)} \rangle_{\mathcal{E}_i}$ to yield the following equality,
\begin{align}
Tr(\mathcal{N}_I) = Tr(T_1 T_2...T_k) = 
\nonumber \\
Tr(\langle \tilde{T}_1^{(\mathbf{e}_1)} \rangle_{\mathcal{E}_1}
\langle \tilde{T}_2^{(\mathbf{e}_2)} \rangle_{\mathcal{E}_2}...
\langle \tilde{T}_k^{(\mathbf{e}_k)} \rangle_{\mathcal{E}_k}).
\end{align}
Allow the symbol $\mathcal{C}_I$ to denote the Cartesian product of the ensembles $\mathcal{E}_1$ through $\mathcal{E}_k$. Then, by linearity, and by the independence of the samples, we have
\begin{align}
Tr(\langle \tilde{T}_1^{(\mathbf{e}_1)} \rangle_{\mathcal{E}_1}
\langle \tilde{T}_2^{(\mathbf{e}_2)} \rangle_{\mathcal{E}_2}...
\langle \tilde{T}_k^{(\mathbf{e}_k)} \rangle_{\mathcal{E}_k}) = 
\nonumber \\
\langle Tr(\tilde{T}_1^{(\mathbf{e}_1)} 
\tilde{T}_2^{(\mathbf{e}_2)}...
\tilde{T}_k^{(\mathbf{e}_k)}) \rangle
_{\mathcal{E}_1\mathcal{E}_2...\mathcal{E}_k} = 
\nonumber \\ 
\langle Tr(\tilde{\mathcal{N}}_{II}^{(\mathbf{c}_I)}) \rangle_{\mathcal{C}_I} = 
\sum_{\mathbf{c}_I \in \mathcal{C}_I} p_{\mathbf{c}_I} Tr(\tilde{\mathcal{N}}_{II}^{(\mathbf{c}_I)}),
\end{align}
where each $\tilde{\mathcal{N}}_{II}^{(\mathbf{c}_I)}$ is the coarse-grained tensor network associated with a particular set of samples $\mathbf{c}_I \in \mathcal{C}_I$ and the application of a single TRG step, and \(p_{\mathbf{c}_I}\) denotes the probability of choosing the collection of samples \(\mathbf{c}_I\). We continue recursively, being careful to note that both the coarse grained tensor networks and the ensembles that allow us to coarse grain them again in an unbiased fashion depend upon our choice of $\mathbf{c}_I$,
\begin{align}
Tr(\tilde{\mathcal{N}}_{II}^{(\mathbf{c}_I)}) =
\langle Tr(\tilde{\mathcal{N}}_{III}^{(\mathbf{c}_{II})}) \rangle_{\mathcal{C}_{II|\mathbf{c}_I}} = 
\nonumber \\
\sum_{\mathbf{c}_{II} \in \mathcal{C}_{II|\mathbf{c}_I}} p_{\mathbf{c}_{II}} Tr(\tilde{\mathcal{N}}_{III}^{(\mathbf{c}_{II})}).
\end{align}
Together then, we find that
\begin{align}
Tr(\mathcal{N}_I) = 
\sum_{\mathbf{c}_I \in \mathcal{C}_I} p_{\mathbf{c}_I} \sum_{\mathbf{c}_{II} \in \mathcal{C}_{II|\mathbf{c}_I}}
p_{\mathbf{c}_{II}} Tr(\tilde{\mathcal{N}}_{III}^{(\mathbf{c}_{II})}) = 
\nonumber \\
\sum_{\mathbf{c}_I \in \mathcal{C}_I} 
\sum_{\mathbf{c}_{II} \in \mathcal{C}_{II|\mathbf{c}_I}} ...
\sum_{\mathbf{c}_{m} \in \mathcal{C}_{m|\mathbf{c}_I \mathbf{c}_{II} ... \mathbf{c}_{m-1}}}
p_{\mathbf{c}_I} p_{\mathbf{c}_{II}} ... p_{\mathbf{c}_m}
Tr(\tilde{\mathcal{N}}_m^{(\mathbf{c}_{m})})
\label{eq:infiniteSum},
\end{align}
where the $\tilde{\mathcal{N}}_m^{(\mathbf{c}_{m})}$, because they contain only a small number of tensors whose bond dimensions have been controlled by the TRG truncations steps, are sufficiently simple that their trace can be computed explicitly.

 We will approximate the sum from Eq. \ref{eq:infiniteSum} by a Monte Carlo sampling. Beginning with $\mathcal{N}_I$, we perform the full singular value decompositions as usual and then choose a subset of singular values to keep according to the proscription in section \ref{sec:tnmc}. Each decomposition is sampled independently, and by completing the coarse graining step as in Ref.~\cite{Levin2007} we generate a coarse grained tensor network $\tilde{\mathcal{N}}_{II}^{(\mathbf{c}_I)}$ with the appropriate probability $p_{\mathbf{c}_I}$. By repeating the same stochastic coarse-graining steps several times we can efficiently sample from the distribution described by Eq. \ref{eq:infiniteSum}. This is a discrete distribution
with a finite number of finite values, therefore it's mean and variance are well defined and we find that
\begin{align}
	\label{eq:infiniteEnsemble}
	\lim_{N \rightarrow \infty} \frac{1}{N} \sum_{i=1}^{i=N} Tr(\tilde{\mathcal{N}}_m^{(i)}) = Tr(\mathcal{N}_I).
\end{align}

\begin{figure*}
  \includegraphics[width=\linewidth]{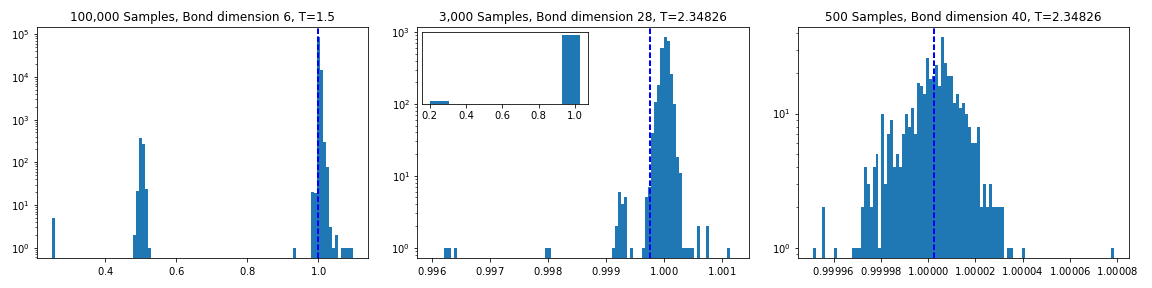}
  \caption{The distribution of results from our stochastic TRG calculations of the partition function at three specific temperature/bond dimension pairs. The dashed line represents the average over all samples and the x axis is scaled so that the exact value of the partition function is one. All data points are shown except one in the middle plot, where there is a single sample only visible in the zoomed-out inset.}
  \label{fig:histograms}
\end{figure*}

We note that it is essential that the samples of different tensors be generated independently, regardless of the symmetries of the physical model. Therefore, the computational time required by this approach scales linearly with the number of tensors in the original network. This growth is comparable to the situation for the deterministic algorithm when the underlying tensor network is not translationally invariant and each tensor must be decomposed separately. It does, however, represent a substantial overhead versus the logarithmic scaling of the non-stochastic approach applied to the case where the underlying system is translationally invariant, and, therefore, every tensor in the network is identical.

\section{\label{sec:ising}Benchmarking with the 2d Classical Ising Model}

In order to benchmark our algorithm we present calculations of the partition function of the 2d classical Ising model at zero field,
\begin{align}
\label{eq:ising_model}
Z = \sum_{\{\sigma\}} e^{\beta\sum_{\langle i,j \rangle}\sigma_i\sigma_j},
\end{align}
both near and far from the critical temperature. All calculations are performed for 128 spins on a periodic lattice using the ITensor library~\footnote{Calculations performed using the ITensor C++ library (version 2.0.7), http://itensor.org/} and compared to an exact summation of the partition function carried out to machine precision. The 8x8 tensor network that represents this partition function is constructed as in Fig. \ref{fig:partition_function_network} and six full renormalization steps are performed, each reducing the size of the network by a factor of two, before the single remaining tensor is traced over.

To understand the characteristics of our Monte Carlo approach to TRG we consider its behavior over a range of bond dimensions and sample sizes. We emphasize again that, in contrast with the deterministic application of TRG, the algorithm discussed in sections \ref{sec:tnmc} and \ref{sec:mcrg} is completely unbiased. That is to say, regardless of the bond dimension, the estimate of the partition function (or any other observable computable by a tensor trace) and the error bars generated for that estimate have no systematic errors and the individual samples are totally uncorrelated. Therefore, we are we are focused on understanding the efficiency of the algorithm with regards to the number of samples needed to attain a desired accuracy as determined by the per-sample variance. 

In Fig.~\ref{fig:histograms} we plot the distribution of individual samples of the partition function generated by our algorithm for several representative choices of bond dimension and temperature. Although the results are unbiased and the samples are uncorrelated regardless of bond dimension, the distribution of samples, the related variances, and overall efficiency of the algorithm will strongly depend on the number of singular values retained at each step. We can gain a qualitative insight into this behavior by examining Fig.~\ref{fig:histograms}, where at small bond dimensions we observe multimodal distributions with one dominant mode and a series of much smaller peaks. This multimodal character implies that a relatively large number of samples would be required to characterize such a distribution and is the source of the large variances that we see at small bond dimensions. Consider the leftmost plot in Fig.~\ref{fig:histograms}, which has more than ninety-nine percent of its probability mass concentrated in the largest mode. If one were to calculate a mean and error bars from a handful of samples it is likely that the smaller peaks would be missed entirely, leading to an overestimate of the mean and an underestimate of the expected error. As the bond dimension increases, these smaller satellite modes are strongly suppressed and the distribution becomes more unimodal, and thus, more amenable to sampling with a restricted number of repetitions.

\begin{table}[h]
    \begin{tabular} {|c|c|c|}
        \hline \vspace{.05cm}
        Bond Dimension & Temperature & Standard Deviation \\ \hline
        6  & 1.5     & 4.171e-02 \\ 
        28 & 2.34826 &  5.017e-04\\ 
        40 & 2.34826 & 1.353e-05 \\ \hline
    \end{tabular} \\ \vskip .2cm
    \caption{The per-sample standard deviation at the three temperature/bond dimension pairs highlighted in Fig. \ref{fig:histograms}. The true value of the partition function has been normalized to one. Notice that the more tightly peaked and unimodal distributions seen at higher bond dimension lead to more accurate individual samples.}
    \label{tab:variances}
\end{table}

The effect that the peculiarities of the underlying distributions have on the quality of the results has been studied extensively in the context of statistical mechanics and in the quantum Monte Carlo community~\cite{Kalos2008MonteMethods, Trail2008}. In some situations one must take care to ensure that the quantities of interest and their variances do not become ill defined because of a poorly posed problem. Tensor network Monte Carlo methods do not suffer from such obstacles because, as noted in section \ref{sec:mcrg}, the summation being sampled actually ranges over a finite (albeit intractably large) number of terms, all of which are themselves finite. However, it is still the case that the efficiency of these techniques depends on having well behaved distributions, as seen in the dramatic drop in per-sample standard deviation (equivalently, the expected error, or \(\sqrt{variance}\)) as the bond dimension is increased that we highlight in table \ref{tab:variances}.

\begin{figure*}[]
  \includegraphics[width=\linewidth]{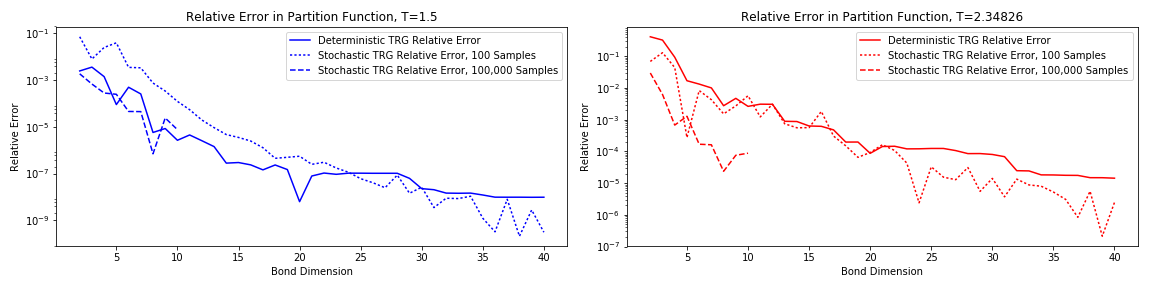}
  \caption{The relative error in the partition function for the deterministic version of TRG alongside the relative error for calculations performed with 100 and 100,000 samples using our stochastic TRG plotted at T=1.5 (left) and T=2.34 (right) over various bond dimensions. We see that the stochastic calculations performed with different numbers of samples follow roughly the same path, except that the curves with more samples are translated downward on a logarithmic scale. This behavior is consistent with the \(\sqrt{1,000}\)-fold decrease in the expectation value of the error guaranteed by Monte Carlo.}
  \label{fig:relativeErrors}
\end{figure*}

When using a deterministic approach to tensor network renormalization one can improve the accuracy of the results only by increasing the bond dimension, whereas a sampling approach provides two ways to accomplish this goal. The first is to increase the number of samples at fixed bond dimension, and the second is to to increase the bond dimension used in each sample. In the limit, as the number of samples, or the bond dimension, goes to infinity, both of these approaches are guaranteed to drive the error to zero. However, while its possible to get the exact partition function or energy at a very small bond dimension using a stochastic approach, it still might be inefficient due to the number of samples needed. To give some perspective on the relative effectiveness of tuning these two parameters we plot a comparison between a deterministic TRG calculation and our MCTRG with a fixed number of samples across a range of bond dimensions. We can see that the deterministic and stochastic versions of TRG become dramatically more accurate with higher bond dimension. While not shown in Fig. \ref{fig:relativeErrors}, the effect of increasing the number of samples is straightforward to understand. Because our samples are calculated independently, we are guaranteed that the expected value of our error will be suppressed inversely proportional to the square root of the number of additional samples we generate. Increasing the number of samples by 100-fold (to 10,000) would therefore result in a relative error 10 times smaller, corresponding to a downward translation of the dotted lines in the logarithmic plots of Fig.~\ref{fig:relativeErrors}.

Interestingly, we see that our stochastic TRG tends to benefit from an increase in bond dimension slightly more than the deterministic version. We speculate that in the low bond dimension regime the deterministic algorithm benefits from fortuitous cancellation of errors while the stochastic approach is beset by the difficulties of sampling from skewed multimodal distributions like those in the left panel of Fig.~\ref{fig:histograms}.
As the bond dimension increases and both of these effects are attenuated the average error per sample in our algorithm drops and we see that we are able to significantly improve on the deterministic results at a given bond dimension by taking a modest number of samples. We will return to these results and consider their usefulness in a wider context in the following section.
Also notable is that in neither case does the increase in accuracy appear monotonic. The jaggedness of the curve for the deterministic results is a well known phenomenon which is frequently encountered in attempts to extrapolate to the infinite bond dimension limit. For enough samples we would expect the stochastic nature of our Monte Carlo version of TRG to smooth out these effects and show a consistent decreases in error with increasing bond dimension but this behavior does not manifest strongly at the sample sizes we have considered.

\section{\label{sec:end} Discussion}
Renormalization group approaches have already proven extremely useful in the quest to take tensor network methods beyond DMRG, but substantial challenges remain. These algorithms can be prohibitively expensive in terms of both time and memory, especially for higher dimensional systems. Furthermore, if the renormalization scheme is not well-suited to the entanglement structure of the tensor network of interest, the bond dimension required for a given accuracy may increase exponentially with the size of the system, or, equivalently, the accuracy may increase very slowly with bond dimension. Extrapolations to the infinite bond dimension limit and estimations of error bounds are challenging to perform~\cite{Corboz2016ImprovedModel}, often rest on unproven heuristics, and can be hampered by a variational bias towards certain states~\cite{Ferris2015}. Finally, these algorithms are labor-intensive to program and optimize even without planning for the parallelizability necessary to make good use of modern computing resources.

Our aim in this paper has been to investigate the feasibility of using the randomized truncation techniques presented in Ref.~\cite{Ferris2015} to alleviate these difficulties. To that end, we implemented a stochastic version of Levin and Nave's tensor renormalization group~\cite{Levin2007}, the simplest of a family of related algorithms, and dissected its performance on the well-studied 2d classical Ising model. We found that, in many cases, an average of a hundred independent samples could outperform a deterministic calculation at an equivalent bond dimension. In other words, considering that the expected value of our error is guaranteed to follow Eq. \ref{eq:shrunkDev}, we found that there exist regimes where the expected error per sample is less than an order of magnitude away from the error of a deterministic calculation.
\begin{align}
\label{eq:shrunkDev}
<error> = \frac{<error\ per\ sample>}{\sqrt{number\ of\ samples}}
\end{align}
We argue that this result is essentially the best that could be hoped for from such a stochastic analogue of TRG. The standard approach of choosing to retain the top $\chi$ singular values is the (locally) optimal one. By randomly choosing a different subset of singular values at each step we expect to do at least a little bit worse with each sample. In exchange for paying this penalty in accuracy per sample, we gain, in addition to unbiased error bars, another parameter besides bond dimension that can be used to systematically improve the accuracy of our calculations. This means that by using parallel computing resources to take more samples, we can arbitrarily suppress the error, controllably approach the exact result, and get an unbiased estimate of the remaining errors. We now turn to highlighting two situations where this additional parameter could be particularly useful, both of which are instances where tools that are most appropriate for one dimensional quantum systems are applied to two dimensional problems.

The asymptotic unsuitability of using a variational tensor network algorithm, like DMRG, designed for a one dimensional quantum system to study a two dimensional one is well understood in terms of entanglement area laws. Less well known is the fact that similar concerns apply to renormalization group approaches as well~\cite{Ferris2013AreaRenormalization}. The 2d classical Ising model that we used in our benchmarking calculations is roughly isomorphic to the imaginary time evolution of a 1d quantum model while, at the same time, the use of Levin and Nave's TRG induces an upper bound on the correlations retained that is similar to what would be obtained when using a tree tensor network as a variational ansatz. 

Given these two facts, it is unsurprising that the accuracy of our deterministic calculations increased so rapidly with bond dimension. However, the use of schemes such as HOTRG, which are still fundamentally limited to a tree-like correlation structure, has recently proved to be a viable approach to treating small two dimensional quantum systems~\cite{Xie2012Coarse-grainingDecomposition}. In such a context, where a renormalization group scheme implicitly tailored to capture the correlations of a one dimensional quantum system is employed to treat a two dimensional one, we would expect a less favorable scaling of accuracy with respect to bond dimension. Under these conditions, the ability of our Monte Carlo approach to reliably suppress error by using a number of samples that grows quadratically with the inverse of the  desired precision could prove vital to the exploration of physics that would otherwise remain out of reach.

To introduce a second example, we note that in the field of variational tensor network calculations the PEPS ansatz has proven to be a powerful tool despite the fact that a naive approach of exactly performing the tensor contractions would require resources that scale exponentially with system size~\cite{Zheng2016StripeModel}. It has been observed that PEPS are capable of representing the ground states of many two dimensional lattice Hamiltonians and that local observables can often be recovered efficiently by approximate methods. Furthermore, rigorous guarantees have been found for these behaviors for a class of gapped Hamiltonians~\cite{Schwarz2017ApproximatingStates}. On the other hand, complexity theoretic arguments~\cite{Schuch2007ComputationalStates} indicate that a large class of quantum states representable by the PEPS ansatz must be challenging to approximately contract, perhaps including the ground states of a variety of gapless Hamiltonians~\cite{Eisert2010iColloquium/iEntropy}. We suggest that the ability of tensor network Monte Carlo calculations to be systematically improved by increasing the number of samples will be especially desirable in situations where the approximate contraction methods currently used for PEPS have difficulties and their accuracy improves in a weak and unreliable fashion with bond dimension. 

Even outside of these regimes where the utility of increasing the bond dimension is limited, a Monte Carlo approach to tensor network renormalization allows for the immediate use of emerging petascale computing infrastructure. While it is likely that some headway could be made in parallelizing existing algorithms and that memory issues could be managed well enough to make somewhat larger bond dimensions accessible, it would be extremely challenging to make full use of high performance computing technology with deterministic tensor network algorithms. By contrast, a stochastic approach is parallelizable with almost no additional effort and will scale dependably with additional computational resources. Our specific finding that a small number of stochastic samples can be competitive in accuracy with a deterministic calculation at the same bond dimension suggests that a wide swath of tensor network techniques could be made more accurate for a reasonable overhead in parallelizable computing resources. 

It is also possible to think of many potential improvements to, and specialized applications of, the approach presented here. For example, even in cases where the underlying system is translationally invariant and one would be reluctant to pay the overhead necessary to do a full stochastic summation we suggest that significant advantages could be had by performing several deterministic coarse-graining steps before switching to a Monte Carlo approach as the number of non-trivial singular values starts to increase. We are also excited about the potential for making use of the unbiased estimates provided by Monte Carlo tensor network contraction as a component of a new PEPS optimization procedure. Furthermore, we are generally hopeful that the unbiased estimate and error bars of a stochastic approach to tensor network renormalization will enhance the interpretability and reliability of calculations performed using these techniques.

\section{Acknowledgements}
This work was supported by
Laboratory Directed Research and Development (LDRD)
funding from Lawrence Berkeley National Laboratory,
provided by the U.S. Department of Energy, Oce of
Science under Contract No. DE-AC02-05CH11231.
CDF was supported by the NSF Graduate Research Fellowship under Grant DGE-1106400 and by the DOE Office of Science Graduate Student Research (SCGSR) program under contract number DESC0014664.
NMT was supported through the Scientific Discovery through
Advanced Computing (SciDAC) program funded by the U.S. Department of
Energy, Office of Science, Advanced Scientific Computing Research and
Basic Energy Sciences.
Computational resources provided by the Extreme
Science and Engineering Discovery Environment (XSEDE), which is
supported by the National Science Foundation Grant No. OCI-1053575,
are gratefully acknowledged.


\begin{thebibliography}{37}%
\makeatletter
\providecommand \@ifxundefined [1]{%
 \@ifx{#1\undefined}
}%
\providecommand \@ifnum [1]{%
 \ifnum #1\expandafter \@firstoftwo
 \else \expandafter \@secondoftwo
 \fi
}%
\providecommand \@ifx [1]{%
 \ifx #1\expandafter \@firstoftwo
 \else \expandafter \@secondoftwo
 \fi
}%
\providecommand \natexlab [1]{#1}%
\providecommand \enquote  [1]{``#1''}%
\providecommand \bibnamefont  [1]{#1}%
\providecommand \bibfnamefont [1]{#1}%
\providecommand \citenamefont [1]{#1}%
\providecommand \href@noop [0]{\@secondoftwo}%
\providecommand \href [0]{\begingroup \@sanitize@url \@href}%
\providecommand \@href[1]{\@@startlink{#1}\@@href}%
\providecommand \@@href[1]{\endgroup#1\@@endlink}%
\providecommand \@sanitize@url [0]{\catcode `\\12\catcode `\$12\catcode
  `\&12\catcode `\#12\catcode `\^12\catcode `\_12\catcode `\%12\relax}%
\providecommand \@@startlink[1]{}%
\providecommand \@@endlink[0]{}%
\providecommand \url  [0]{\begingroup\@sanitize@url \@url }%
\providecommand \@url [1]{\endgroup\@href {#1}{\urlprefix }}%
\providecommand \urlprefix  [0]{URL }%
\providecommand \Eprint [0]{\href }%
\providecommand \doibase [0]{http://dx.doi.org/}%
\providecommand \selectlanguage [0]{\@gobble}%
\providecommand \bibinfo  [0]{\@secondoftwo}%
\providecommand \bibfield  [0]{\@secondoftwo}%
\providecommand \translation [1]{[#1]}%
\providecommand \BibitemOpen [0]{}%
\providecommand \bibitemStop [0]{}%
\providecommand \bibitemNoStop [0]{.\EOS\space}%
\providecommand \EOS [0]{\spacefactor3000\relax}%
\providecommand \BibitemShut  [1]{\csname bibitem#1\endcsname}%
\let\auto@bib@innerbib\@empty
\bibitem [{\citenamefont {White}(1992)}]{White1992}%
  \BibitemOpen
  \bibfield  {author} {\bibinfo {author} {\bibfnamefont {S.~R.}\ \bibnamefont
  {White}},\ }\href {\doibase 10.1103/PhysRevLett.69.2863} {\bibfield
  {journal} {\bibinfo  {journal} {Phys. Rev. Lett.}\ }\textbf {\bibinfo
  {volume} {69}},\ \bibinfo {pages} {2863} (\bibinfo {year}
  {1992})}\BibitemShut {NoStop}%
\bibitem [{\citenamefont {Schollw{\"o}ck}(2011)}]{dmrg2011}%
  \BibitemOpen
  \bibfield  {author} {\bibinfo {author} {\bibfnamefont {U.}~\bibnamefont
  {Schollw{\"o}ck}},\ }\href {\doibase 10.1016/j.aop.2010.09.012} {\bibfield
  {journal} {\bibinfo  {journal} {Ann.~Phys.}\ }\textbf {\bibinfo {volume}
  {326}},\ \bibinfo {pages} {96 } (\bibinfo {year} {2011})}\BibitemShut
  {NoStop}%
\bibitem [{\citenamefont {White}(1993)}]{white1993}%
  \BibitemOpen
  \bibfield  {author} {\bibinfo {author} {\bibfnamefont {S.~R.}\ \bibnamefont
  {White}},\ }\href {\doibase 10.1103/PhysRevB.48.10345} {\bibfield  {journal}
  {\bibinfo  {journal} {Phys. Rev. B}\ }\textbf {\bibinfo {volume} {48}},\
  \bibinfo {pages} {10345} (\bibinfo {year} {1993})}\BibitemShut {NoStop}%
\bibitem [{\citenamefont {White}\ and\ \citenamefont
  {Martin}(1999)}]{white1999}%
  \BibitemOpen
  \bibfield  {author} {\bibinfo {author} {\bibfnamefont {S.~R.}\ \bibnamefont
  {White}}\ and\ \bibinfo {author} {\bibfnamefont {R.~L.}\ \bibnamefont
  {Martin}},\ }\href {\doibase 10.1063/1.478295} {\bibfield  {journal}
  {\bibinfo  {journal} {J. Chem. Phys.}\ }\textbf {\bibinfo {volume} {110}},\
  \bibinfo {pages} {4127} (\bibinfo {year} {1999})}\BibitemShut {NoStop}%
\bibitem [{\citenamefont {Kurashige}\ and\ \citenamefont
  {Yanai}(2009)}]{yanai2009}%
  \BibitemOpen
  \bibfield  {author} {\bibinfo {author} {\bibfnamefont {Y.}~\bibnamefont
  {Kurashige}}\ and\ \bibinfo {author} {\bibfnamefont {T.}~\bibnamefont
  {Yanai}},\ }\href
  {http://scitation.aip.org/content/aip/journal/jcp/130/23/10.1063/1.3152576}
  {\bibfield  {journal} {\bibinfo  {journal} {J. Chem. Phys.}\ }\textbf
  {\bibinfo {volume} {130}},\ \bibinfo {eid} {234114} (\bibinfo {year}
  {2009})}\BibitemShut {NoStop}%
\bibitem [{\citenamefont {Stoudenmire}\ and\ \citenamefont
  {White}(2012)}]{stoudenmire2012}%
  \BibitemOpen
  \bibfield  {author} {\bibinfo {author} {\bibfnamefont {E.}~\bibnamefont
  {Stoudenmire}}\ and\ \bibinfo {author} {\bibfnamefont {S.~R.}\ \bibnamefont
  {White}},\ }\href@noop {} {\bibfield  {journal} {\bibinfo  {journal} {Annu.
  Rev. Conden. Ma. P.}\ }\textbf {\bibinfo {volume} {3}},\ \bibinfo {pages}
  {111} (\bibinfo {year} {2012})}\BibitemShut {NoStop}%
\bibitem [{\citenamefont {Changlani}\ \emph {et~al.}(2016)\citenamefont
  {Changlani}, \citenamefont {Tubman},\ and\ \citenamefont
  {Hughes}}]{Changlani2016}%
  \BibitemOpen
  \bibfield  {author} {\bibinfo {author} {\bibfnamefont {H.~J.}\ \bibnamefont
  {Changlani}}, \bibinfo {author} {\bibfnamefont {N.~M.}\ \bibnamefont
  {Tubman}}, \ and\ \bibinfo {author} {\bibfnamefont {T.~L.}\ \bibnamefont
  {Hughes}},\ }\href {http://dx.doi.org/10.1038/srep31897} {\bibfield
  {journal} {\bibinfo  {journal} {Sci. Rep.}\ }\textbf {\bibinfo {volume}
  {6}},\ \bibinfo {pages} {31897} (\bibinfo {year} {2016})}\BibitemShut
  {NoStop}%
\bibitem [{\citenamefont {Olivares-Amaya}\ \emph {et~al.}(2015)\citenamefont
  {Olivares-Amaya}, \citenamefont {Hu}, \citenamefont {Nakatani}, \citenamefont
  {Sharma}, \citenamefont {Yang},\ and\ \citenamefont {Chan}}]{amaya2015}%
  \BibitemOpen
  \bibfield  {author} {\bibinfo {author} {\bibfnamefont {R.}~\bibnamefont
  {Olivares-Amaya}}, \bibinfo {author} {\bibfnamefont {W.}~\bibnamefont {Hu}},
  \bibinfo {author} {\bibfnamefont {N.}~\bibnamefont {Nakatani}}, \bibinfo
  {author} {\bibfnamefont {S.}~\bibnamefont {Sharma}}, \bibinfo {author}
  {\bibfnamefont {J.}~\bibnamefont {Yang}}, \ and\ \bibinfo {author}
  {\bibfnamefont {G.~K.-L.}\ \bibnamefont {Chan}},\ }\href
  {http://scitation.aip.org/content/aip/journal/jcp/142/3/10.1063/1.4905329}
  {\bibfield  {journal} {\bibinfo  {journal} {J. Chem. Phys.}\ }\textbf
  {\bibinfo {volume} {142}},\ \bibinfo {eid} {034102} (\bibinfo {year}
  {2015})}\BibitemShut {NoStop}%
\bibitem [{\citenamefont {Dukelsky}\ \emph {et~al.}(1998)\citenamefont
  {Dukelsky}, \citenamefont {Mart{\'{i}}n-Delgado}, \citenamefont {Nishino},
  \citenamefont {Sierra}, \citenamefont {Martin-Delgado}, \citenamefont
  {Nishino},\ and\ \citenamefont {Sierra}}]{Dukelsky1998}%
  \BibitemOpen
  \bibfield  {author} {\bibinfo {author} {\bibfnamefont {J.}~\bibnamefont
  {Dukelsky}}, \bibinfo {author} {\bibfnamefont {M.~A.}\ \bibnamefont
  {Mart{\'{i}}n-Delgado}}, \bibinfo {author} {\bibfnamefont {T.}~\bibnamefont
  {Nishino}}, \bibinfo {author} {\bibfnamefont {G.}~\bibnamefont {Sierra}},
  \bibinfo {author} {\bibfnamefont {M.~A.}\ \bibnamefont {Martin-Delgado}},
  \bibinfo {author} {\bibfnamefont {T.}~\bibnamefont {Nishino}}, \ and\
  \bibinfo {author} {\bibfnamefont {G.}~\bibnamefont {Sierra}},\ }\href
  {\doibase 10.1209/epl/i1998-00381-x} {\bibfield  {journal} {\bibinfo
  {journal} {Europhys. Lett.}\ }\textbf {\bibinfo {volume} {43}},\ \bibinfo
  {pages} {457} (\bibinfo {year} {1998})}\BibitemShut {NoStop}%
\bibitem [{\citenamefont {Verstraete}\ \emph {et~al.}(2008)\citenamefont
  {Verstraete}, \citenamefont {Murg},\ and\ \citenamefont
  {Cirac}}]{Verstraete2008}%
  \BibitemOpen
  \bibfield  {author} {\bibinfo {author} {\bibfnamefont {F.}~\bibnamefont
  {Verstraete}}, \bibinfo {author} {\bibfnamefont {V.}~\bibnamefont {Murg}}, \
  and\ \bibinfo {author} {\bibfnamefont {J.}~\bibnamefont {Cirac}},\ }\href
  {\doibase 10.1080/14789940801912366} {\bibfield  {journal} {\bibinfo
  {journal} {Adv. Phys.}\ }\textbf {\bibinfo {volume} {57}},\ \bibinfo {pages}
  {143} (\bibinfo {year} {2008})}\BibitemShut {NoStop}%
\bibitem [{\citenamefont {Vidal}(2008)}]{Vidal2008}%
  \BibitemOpen
  \bibfield  {author} {\bibinfo {author} {\bibfnamefont {G.}~\bibnamefont
  {Vidal}},\ }\href {\doibase 10.1103/PhysRevLett.101.110501} {\bibfield
  {journal} {\bibinfo  {journal} {Phys. Rev. Lett.}\ }\textbf {\bibinfo
  {volume} {101}},\ \bibinfo {pages} {110501} (\bibinfo {year}
  {2008})}\BibitemShut {NoStop}%
\bibitem [{\citenamefont {Levin}\ and\ \citenamefont {Nave}(2007)}]{Levin2007}%
  \BibitemOpen
  \bibfield  {author} {\bibinfo {author} {\bibfnamefont {M.}~\bibnamefont
  {Levin}}\ and\ \bibinfo {author} {\bibfnamefont {C.~P.}\ \bibnamefont
  {Nave}},\ }\href {\doibase 10.1103/PhysRevLett.99.120601} {\bibfield
  {journal} {\bibinfo  {journal} {Phys. Rev. Lett.}\ }\textbf {\bibinfo
  {volume} {99}},\ \bibinfo {pages} {120601} (\bibinfo {year}
  {2007})}\BibitemShut {NoStop}%
\bibitem [{\citenamefont {Xie}\ \emph {et~al.}(2012)\citenamefont {Xie},
  \citenamefont {Chen}, \citenamefont {Qin}, \citenamefont {Zhu}, \citenamefont
  {Yang},\ and\ \citenamefont {Xiang}}]{Xie2012Coarse-grainingDecomposition}%
  \BibitemOpen
  \bibfield  {author} {\bibinfo {author} {\bibfnamefont {Z.~Y.}\ \bibnamefont
  {Xie}}, \bibinfo {author} {\bibfnamefont {J.}~\bibnamefont {Chen}}, \bibinfo
  {author} {\bibfnamefont {M.~P.}\ \bibnamefont {Qin}}, \bibinfo {author}
  {\bibfnamefont {J.~W.}\ \bibnamefont {Zhu}}, \bibinfo {author} {\bibfnamefont
  {L.~P.}\ \bibnamefont {Yang}}, \ and\ \bibinfo {author} {\bibfnamefont
  {T.}~\bibnamefont {Xiang}},\ }\href {\doibase 10.1103/PhysRevB.86.045139}
  {\bibfield  {journal} {\bibinfo  {journal} {Phys. Rev. B}\ }\textbf {\bibinfo
  {volume} {86}},\ \bibinfo {pages} {045139} (\bibinfo {year}
  {2012})}\BibitemShut {NoStop}%
\bibitem [{\citenamefont {Evenbly}\ and\ \citenamefont
  {Vidal}(2015)}]{Evenbly2015}%
  \BibitemOpen
  \bibfield  {author} {\bibinfo {author} {\bibfnamefont {G.}~\bibnamefont
  {Evenbly}}\ and\ \bibinfo {author} {\bibfnamefont {G.}~\bibnamefont
  {Vidal}},\ }\href {\doibase 10.1103/PhysRevLett.115.180405} {\bibfield
  {journal} {\bibinfo  {journal} {Phys. Rev. Lett.}\ }\textbf {\bibinfo
  {volume} {115}},\ \bibinfo {pages} {180405} (\bibinfo {year}
  {2015})}\BibitemShut {NoStop}%
\bibitem [{\citenamefont {Nishino}\ and\ \citenamefont
  {Okunishi}(1996)}]{Nishino1996CornerMethod}%
  \BibitemOpen
  \bibfield  {author} {\bibinfo {author} {\bibfnamefont {T.}~\bibnamefont
  {Nishino}}\ and\ \bibinfo {author} {\bibfnamefont {K.}~\bibnamefont
  {Okunishi}},\ }\href {\doibase 10.1143/JPSJ.65.891} {\bibfield  {journal}
  {\bibinfo  {journal} {J. Phys. Soc Jpn.}\ }\textbf {\bibinfo {volume} {65}},\
  \bibinfo {pages} {891} (\bibinfo {year} {1996})}\BibitemShut {NoStop}%
\bibitem [{\citenamefont {Corboz}\ \emph {et~al.}(2010)\citenamefont {Corboz},
  \citenamefont {Or{\'{u}}s}, \citenamefont {Bauer},\ and\ \citenamefont
  {Vidal}}]{Corboz2010SimulationStates}%
  \BibitemOpen
  \bibfield  {author} {\bibinfo {author} {\bibfnamefont {P.}~\bibnamefont
  {Corboz}}, \bibinfo {author} {\bibfnamefont {R.}~\bibnamefont {Or{\'{u}}s}},
  \bibinfo {author} {\bibfnamefont {B.}~\bibnamefont {Bauer}}, \ and\ \bibinfo
  {author} {\bibfnamefont {G.}~\bibnamefont {Vidal}},\ }\href {\doibase
  10.1103/PhysRevB.81.165104} {\bibfield  {journal} {\bibinfo  {journal} {Phys.
  Rev. B}\ }\textbf {\bibinfo {volume} {81}},\ \bibinfo {pages} {165104}
  (\bibinfo {year} {2010})}\BibitemShut {NoStop}%
\bibitem [{\citenamefont {Corboz}(2016)}]{Corboz2016ImprovedModel}%
  \BibitemOpen
  \bibfield  {author} {\bibinfo {author} {\bibfnamefont {P.}~\bibnamefont
  {Corboz}},\ }\href {\doibase 10.1103/PhysRevB.93.045116} {\bibfield
  {journal} {\bibinfo  {journal} {Phys. Rev. B}\ }\textbf {\bibinfo {volume}
  {93}},\ \bibinfo {pages} {045116} (\bibinfo {year} {2016})}\BibitemShut
  {NoStop}%
\bibitem [{\citenamefont {Zheng}\ \emph {et~al.}(2016)\citenamefont {Zheng},
  \citenamefont {Chung}, \citenamefont {Corboz}, \citenamefont {Ehlers},
  \citenamefont {Qin}, \citenamefont {Noack}, \citenamefont {Shi},
  \citenamefont {White}, \citenamefont {Zhang},\ and\ \citenamefont
  {Chan}}]{Zheng2016StripeModel}%
  \BibitemOpen
  \bibfield  {author} {\bibinfo {author} {\bibfnamefont {B.-X.}\ \bibnamefont
  {Zheng}}, \bibinfo {author} {\bibfnamefont {C.-M.}\ \bibnamefont {Chung}},
  \bibinfo {author} {\bibfnamefont {P.}~\bibnamefont {Corboz}}, \bibinfo
  {author} {\bibfnamefont {G.}~\bibnamefont {Ehlers}}, \bibinfo {author}
  {\bibfnamefont {M.-P.}\ \bibnamefont {Qin}}, \bibinfo {author} {\bibfnamefont
  {R.~M.}\ \bibnamefont {Noack}}, \bibinfo {author} {\bibfnamefont
  {H.}~\bibnamefont {Shi}}, \bibinfo {author} {\bibfnamefont {S.~R.}\
  \bibnamefont {White}}, \bibinfo {author} {\bibfnamefont {S.}~\bibnamefont
  {Zhang}}, \ and\ \bibinfo {author} {\bibfnamefont {G.~K.-L.}\ \bibnamefont
  {Chan}},\ }\href {http://arxiv.org/abs/1701.00054} {\  (\bibinfo {year}
  {2016})},\ \Eprint {http://arxiv.org/abs/1701.00054} {arXiv:1701.00054}
  \BibitemShut {NoStop}%
\bibitem [{\citenamefont {{Bruognolo}}\ \emph {et~al.}(2017)\citenamefont
  {{Bruognolo}}, \citenamefont {{Zhu}}, \citenamefont {{White}},\ and\
  \citenamefont {{Miles Stoudenmire}}}]{bruognolo2017}%
  \BibitemOpen
  \bibfield  {author} {\bibinfo {author} {\bibfnamefont {B.}~\bibnamefont
  {{Bruognolo}}}, \bibinfo {author} {\bibfnamefont {Z.}~\bibnamefont {{Zhu}}},
  \bibinfo {author} {\bibfnamefont {S.~R.}\ \bibnamefont {{White}}}, \ and\
  \bibinfo {author} {\bibfnamefont {E.}~\bibnamefont {{Miles Stoudenmire}}},\
  }\href@noop {} {\  (\bibinfo {year} {2017})},\ \Eprint
  {http://arxiv.org/abs/1705.05578} {arXiv:1705.05578} \BibitemShut {NoStop}%
\bibitem [{\citenamefont {Yan}\ \emph {et~al.}(2011)\citenamefont {Yan},
  \citenamefont {Huse},\ and\ \citenamefont {White}}]{simeng2010}%
  \BibitemOpen
  \bibfield  {author} {\bibinfo {author} {\bibfnamefont {S.}~\bibnamefont
  {Yan}}, \bibinfo {author} {\bibfnamefont {D.~A.}\ \bibnamefont {Huse}}, \
  and\ \bibinfo {author} {\bibfnamefont {S.~R.}\ \bibnamefont {White}},\
  }\href@noop {} {\bibfield  {journal} {\bibinfo  {journal} {Science}\ }\textbf
  {\bibinfo {volume} {332}},\ \bibinfo {pages} {1173} (\bibinfo {year}
  {2011})}\BibitemShut {NoStop}%
\bibitem [{\citenamefont {Or{\'{u}}s}(2012)}]{Orus2012ExploringSystems}%
  \BibitemOpen
  \bibfield  {author} {\bibinfo {author} {\bibfnamefont {R.}~\bibnamefont
  {Or{\'{u}}s}},\ }\href {\doibase 10.1103/PhysRevB.85.205117} {\bibfield
  {journal} {\bibinfo  {journal} {Phys. Rev. B}\ }\textbf {\bibinfo {volume}
  {85}},\ \bibinfo {pages} {205117} (\bibinfo {year} {2012})}\BibitemShut
  {NoStop}%
\bibitem [{\citenamefont {Schwarz}\ \emph {et~al.}(2017)\citenamefont
  {Schwarz}, \citenamefont {Buerschaper},\ and\ \citenamefont
  {Eisert}}]{Schwarz2017ApproximatingStates}%
  \BibitemOpen
  \bibfield  {author} {\bibinfo {author} {\bibfnamefont {M.}~\bibnamefont
  {Schwarz}}, \bibinfo {author} {\bibfnamefont {O.}~\bibnamefont
  {Buerschaper}}, \ and\ \bibinfo {author} {\bibfnamefont {J.}~\bibnamefont
  {Eisert}},\ }\href {\doibase 10.1103/PhysRevA.95.060102} {\bibfield
  {journal} {\bibinfo  {journal} {Phys. Rev. A}\ }\textbf {\bibinfo {volume}
  {95}},\ \bibinfo {pages} {060102} (\bibinfo {year} {2017})}\BibitemShut
  {NoStop}%
\bibitem [{\citenamefont {Chan}(2004)}]{Chan2004AnCalculations}%
  \BibitemOpen
  \bibfield  {author} {\bibinfo {author} {\bibfnamefont {G.~K.-L.}\
  \bibnamefont {Chan}},\ }\href {\doibase 10.1063/1.1638734} {\bibfield
  {journal} {\bibinfo  {journal} {J. Chem. Phys.}\ }\textbf {\bibinfo {volume}
  {120}},\ \bibinfo {pages} {3172} (\bibinfo {year} {2004})}\BibitemShut
  {NoStop}%
\bibitem [{\citenamefont {Stoudenmire}\ and\ \citenamefont
  {White}(2013)}]{stoudenmire2013}%
  \BibitemOpen
  \bibfield  {author} {\bibinfo {author} {\bibfnamefont {E.~M.}\ \bibnamefont
  {Stoudenmire}}\ and\ \bibinfo {author} {\bibfnamefont {S.~R.}\ \bibnamefont
  {White}},\ }\href {\doibase 10.1103/PhysRevB.87.155137} {\bibfield  {journal}
  {\bibinfo  {journal} {Phys. Rev. B}\ }\textbf {\bibinfo {volume} {87}},\
  \bibinfo {pages} {155137} (\bibinfo {year} {2013})}\BibitemShut {NoStop}%
\bibitem [{\citenamefont {Needs}\ \emph {et~al.}(2010)\citenamefont {Needs},
  \citenamefont {Towler}, \citenamefont {Drummond},\ and\ \citenamefont
  {R\'ios}}]{needs2010}%
  \BibitemOpen
  \bibfield  {author} {\bibinfo {author} {\bibfnamefont {R.~J.}\ \bibnamefont
  {Needs}}, \bibinfo {author} {\bibfnamefont {M.~D.}\ \bibnamefont {Towler}},
  \bibinfo {author} {\bibfnamefont {N.~D.}\ \bibnamefont {Drummond}}, \ and\
  \bibinfo {author} {\bibfnamefont {P.~L.}\ \bibnamefont {R\'ios}},\ }\href
  {http://stacks.iop.org/0953-8984/22/i=2/a=023201} {\bibfield  {journal}
  {\bibinfo  {journal} {J. Phys-Condens. Mat.}\ }\textbf {\bibinfo {volume}
  {22}},\ \bibinfo {pages} {023201} (\bibinfo {year} {2010})}\BibitemShut
  {NoStop}%
\bibitem [{\citenamefont {Tubman}\ \emph {et~al.}(2011)\citenamefont {Tubman},
  \citenamefont {DuBois}, \citenamefont {Hood},\ and\ \citenamefont
  {Alder}}]{tubman2011}%
  \BibitemOpen
  \bibfield  {author} {\bibinfo {author} {\bibfnamefont {N.~M.}\ \bibnamefont
  {Tubman}}, \bibinfo {author} {\bibfnamefont {J.~L.}\ \bibnamefont {DuBois}},
  \bibinfo {author} {\bibfnamefont {R.~Q.}\ \bibnamefont {Hood}}, \ and\
  \bibinfo {author} {\bibfnamefont {B.~J.}\ \bibnamefont {Alder}},\ }\href@noop
  {} {\bibfield  {journal} {\bibinfo  {journal} {J. Chem. Phys.}\ }\textbf
  {\bibinfo {volume} {135}},\ \bibinfo {pages} {184109} (\bibinfo {year}
  {2011})}\BibitemShut {NoStop}%
\bibitem [{\citenamefont {Tubman}\ \emph {et~al.}(2014)\citenamefont {Tubman},
  \citenamefont {Kyl\"anp\"a\"a}, \citenamefont {Hammes-Schiffer},\ and\
  \citenamefont {Ceperley}}]{tubman2014}%
  \BibitemOpen
  \bibfield  {author} {\bibinfo {author} {\bibfnamefont {N.~M.}\ \bibnamefont
  {Tubman}}, \bibinfo {author} {\bibfnamefont {I.}~\bibnamefont
  {Kyl\"anp\"a\"a}}, \bibinfo {author} {\bibfnamefont {S.}~\bibnamefont
  {Hammes-Schiffer}}, \ and\ \bibinfo {author} {\bibfnamefont {D.~M.}\
  \bibnamefont {Ceperley}},\ }\href {\doibase 10.1103/PhysRevA.90.042507}
  {\bibfield  {journal} {\bibinfo  {journal} {Phys. Rev. A}\ }\textbf {\bibinfo
  {volume} {90}},\ \bibinfo {pages} {042507} (\bibinfo {year}
  {2014})}\BibitemShut {NoStop}%
\bibitem [{\citenamefont {Brown}\ \emph {et~al.}(2013)\citenamefont {Brown},
  \citenamefont {Clark}, \citenamefont {DuBois},\ and\ \citenamefont
  {Ceperley}}]{brown2013}%
  \BibitemOpen
  \bibfield  {author} {\bibinfo {author} {\bibfnamefont {E.~W.}\ \bibnamefont
  {Brown}}, \bibinfo {author} {\bibfnamefont {B.~K.}\ \bibnamefont {Clark}},
  \bibinfo {author} {\bibfnamefont {J.~L.}\ \bibnamefont {DuBois}}, \ and\
  \bibinfo {author} {\bibfnamefont {D.~M.}\ \bibnamefont {Ceperley}},\ }\href
  {\doibase 10.1103/PhysRevLett.110.146405} {\bibfield  {journal} {\bibinfo
  {journal} {Phys. Rev. Lett.}\ }\textbf {\bibinfo {volume} {110}},\ \bibinfo
  {pages} {146405} (\bibinfo {year} {2013})}\BibitemShut {NoStop}%
\bibitem [{\citenamefont {Zhang}\ and\ \citenamefont
  {Krakauer}(2003)}]{zhang2003}%
  \BibitemOpen
  \bibfield  {author} {\bibinfo {author} {\bibfnamefont {S.}~\bibnamefont
  {Zhang}}\ and\ \bibinfo {author} {\bibfnamefont {H.}~\bibnamefont
  {Krakauer}},\ }\href {\doibase 10.1103/PhysRevLett.90.136401} {\bibfield
  {journal} {\bibinfo  {journal} {Phys. Rev. Lett.}\ }\textbf {\bibinfo
  {volume} {90}},\ \bibinfo {pages} {136401} (\bibinfo {year}
  {2003})}\BibitemShut {NoStop}%
\bibitem [{\citenamefont {Ferris}(2015)}]{Ferris2015}%
  \BibitemOpen
  \bibfield  {author} {\bibinfo {author} {\bibfnamefont {A.~J.}\ \bibnamefont
  {Ferris}},\ }\href {http://arxiv.org/abs/1507.0767} {\  (\bibinfo {year}
  {2015})},\ \Eprint {http://arxiv.org/abs/1507.0767} {arXiv:1507.0767}
  \BibitemShut {NoStop}%
\bibitem [{\citenamefont {Eckart}\ and\ \citenamefont
  {Young}(1936)}]{Eckart1936TheRank}%
  \BibitemOpen
  \bibfield  {author} {\bibinfo {author} {\bibfnamefont {C.}~\bibnamefont
  {Eckart}}\ and\ \bibinfo {author} {\bibfnamefont {G.}~\bibnamefont {Young}},\
  }\href {\doibase 10.1007/BF02288367} {\bibfield  {journal} {\bibinfo
  {journal} {Psychometrika}\ }\textbf {\bibinfo {volume} {1}},\ \bibinfo
  {pages} {211} (\bibinfo {year} {1936})}\BibitemShut {NoStop}%
\bibitem [{Note1()}]{Note1}%
  \BibitemOpen
  \bibinfo {note} {Calculations performed using the ITensor C++ library
  (version 2.0.7), http://itensor.org/}\BibitemShut {NoStop}%
\bibitem [{\citenamefont {Kalos}\ and\ \citenamefont
  {Whitlock}(2008)}]{Kalos2008MonteMethods}%
  \BibitemOpen
  \bibfield  {author} {\bibinfo {author} {\bibfnamefont {M.~H.}\ \bibnamefont
  {Kalos}}\ and\ \bibinfo {author} {\bibfnamefont {P.~A.}\ \bibnamefont
  {Whitlock}},\ }\href {\doibase 10.1002/9783527626212} {\emph {\bibinfo
  {title} {{Monte Carlo Methods}}}}\ (\bibinfo  {publisher} {Wiley-VCH Verlag
  GmbH {\&} Co. KGaA},\ \bibinfo {address} {Weinheim, Germany},\ \bibinfo
  {year} {2008})\BibitemShut {NoStop}%
\bibitem [{\citenamefont {Trail}(2008)}]{Trail2008}%
  \BibitemOpen
  \bibfield  {author} {\bibinfo {author} {\bibfnamefont {J.~R.}\ \bibnamefont
  {Trail}},\ }\href {\doibase 10.1103/PhysRevE.77.016703} {\bibfield  {journal}
  {\bibinfo  {journal} {Phys. Rev. E}\ }\textbf {\bibinfo {volume} {77}},\
  \bibinfo {pages} {016703} (\bibinfo {year} {2008})}\BibitemShut {NoStop}%
\bibitem [{\citenamefont {Ferris}(2013)}]{Ferris2013AreaRenormalization}%
  \BibitemOpen
  \bibfield  {author} {\bibinfo {author} {\bibfnamefont {A.~J.}\ \bibnamefont
  {Ferris}},\ }\href {\doibase 10.1103/PhysRevB.87.125139} {\bibfield
  {journal} {\bibinfo  {journal} {Phys. Rev. B}\ }\textbf {\bibinfo {volume}
  {87}},\ \bibinfo {pages} {125139} (\bibinfo {year} {2013})}\BibitemShut
  {NoStop}%
\bibitem [{\citenamefont {Schuch}\ \emph {et~al.}(2007)\citenamefont {Schuch},
  \citenamefont {Wolf}, \citenamefont {Verstraete},\ and\ \citenamefont
  {Cirac}}]{Schuch2007ComputationalStates}%
  \BibitemOpen
  \bibfield  {author} {\bibinfo {author} {\bibfnamefont {N.}~\bibnamefont
  {Schuch}}, \bibinfo {author} {\bibfnamefont {M.~M.}\ \bibnamefont {Wolf}},
  \bibinfo {author} {\bibfnamefont {F.}~\bibnamefont {Verstraete}}, \ and\
  \bibinfo {author} {\bibfnamefont {J.~I.}\ \bibnamefont {Cirac}},\ }\href
  {\doibase 10.1103/PhysRevLett.98.140506} {\bibfield  {journal} {\bibinfo
  {journal} {Phys. Rev. Lett.}\ }\textbf {\bibinfo {volume} {98}},\ \bibinfo
  {pages} {140506} (\bibinfo {year} {2007})}\BibitemShut {NoStop}%
\bibitem [{\citenamefont {Eisert}\ \emph {et~al.}(2010)\citenamefont {Eisert},
  \citenamefont {Cramer},\ and\ \citenamefont
  {Plenio}}]{Eisert2010iColloquium/iEntropy}%
  \BibitemOpen
  \bibfield  {author} {\bibinfo {author} {\bibfnamefont {J.}~\bibnamefont
  {Eisert}}, \bibinfo {author} {\bibfnamefont {M.}~\bibnamefont {Cramer}}, \
  and\ \bibinfo {author} {\bibfnamefont {M.~B.}\ \bibnamefont {Plenio}},\
  }\href {\doibase 10.1103/RevModPhys.82.277} {\bibfield  {journal} {\bibinfo
  {journal} {Rev. Mod. Phys.}\ }\textbf {\bibinfo {volume} {82}},\ \bibinfo
  {pages} {277} (\bibinfo {year} {2010})}\BibitemShut {NoStop}%
\end{thebibliography}
%
\end{document}